\documentclass[11pt,twoside]{article}

\usepackage{asp2006}
\usepackage{epsf}
\usepackage{psfig}
\usepackage{lscape}
\usepackage{graphicx}

\begin{document}
\title{Advances in Reverberation Mapping} 
\author{Shai Kaspi} 
\affil{School of Physics \& Astronomy and the Wise Observatory,
Tel-Aviv University Tel-Aviv 69978, Israel \\ 
Physics Department, Technion, Haifa 32000, Israel}

\begin{abstract} 
This contribution briefly reviews the reverberation mapping technique
which leads to determination of black hole masses. I focus on the
emerging relation between the broad-line region size and the active
galactic nucleus (AGN) luminosity, and on an overview of recent results
of reverberation mapping studies which are starting to cover the full
AGN luminosity range. Preliminary results and time lag determination
from a reverberation mapping program of high-luminosity quasars are
also presented.
\end{abstract}


\section{Introduction} 

While the physical origin of Active Galactic Nuclei (AGNs) continuum
variability is still unclear, it is possible to use the ``reverberation
mapping'' technique to study the geometry and kinematics of the gas
in the Broad Line Region (BLR), and to deduce the mass of the central
black hole (BH) in the center of the AGN. The technique is based on
the response of the ambient gas to changes in the central continuum
source. Such response was first used to account for the observations
of the apparent expansion of Nova Persei in 1901 (Couderc 1939) and
has been proposed to explain the light curves of Type I supernovae
(Morrison \& Sartori 1969). The method was first suggested to be
used in the analysis of AGN light curves by Bahcall, Kozlovsky, \&
Salpeter (1972), who calculated the response of the line intensity
in a spherical distribution of gas. Blandford \& McKee (1982) were
the first to coin the term ``reverberation mapping'' and put it into
mathematical formalism with the fundamental equation that relates
the emission-line and continuum light curves, $L(v,t)$ and $C(t)$:
\begin{equation}
L(v,t)=\int\Psi(v,t-\tau)C(\tau)d\tau,
\end{equation}
where $v$ is the velocity field of the BLR (which manifest itself in
the emission-line profile), and $\Psi(v,\tau)$ is defined from this
equation as the transfer function, which holds in it the information
about the geometry and kinematics of the BLR. The latter was studied
by several authors who showed how $\Psi(v,\tau)$ can be derived from
the observed continuum and line light curves and how $\Psi(v,\tau)$
will changed for different geometries and kinematics of the BLR (e.g.,
Welsh \& Horne 1991; Perez et al. 1992; Horne et al. 2004).

In practice, in order to get the line and continuum light curves,
the AGN needs to be monitored frequently over a period of time
(from days to weeks depending on the AGN luminosity and variability
characteristics). First attempts to carry out AGN reverberation
mapping used poorly sampled light curves and low resolution spectra
(e.g., Peterson 1988 and references therein), thus leading to a
collapse of the two-dimension transfer function, $\Psi(v,\tau)$,
into one-dimension transfer function $\Psi(\tau)$. In fact this is
further collapsed into only one parameter: the time lag between
the line light curve and the continuum light curve. The time lag
is defined as the centroid of the cross correlation function (CCF)
between the continuum and line light curves:
\begin{equation}
F_{CCF}(\tau)=(N\sigma_C\sigma_L)^{-1}\sum_{t}C(t)L(t+\tau),
\end{equation}
where N is the number of points used in the sum for the lag $\tau$,
$\sigma_C$ and $\sigma_L$ are the rms of the light curves, and the
light curves have zero mean. The centroid of this CCF is taken to be a
measure for the size of the BLR, denotes as $R_{BLR}$. Once $R_{BLR}$
is found from reverberation mapping the mass of the BH in the center
of the AGN can be estimated using:
\begin{equation}
M_{BH}=fG^{-1}R_{BLR}V^2,
\end{equation}
where $V$ is a measure of the BLR clouds' velocity, and $f$ is a
dimensionless factor that depends of the geometry and kinematics of the
BLR. The mass---luminosity relation is further discussed in Peterson's
contribution to these proceedings and will not be discussed here.

Over the past two decades many monitoring campaigns were carried
out and enabled the measurement of $R_{BLR}$ in about three
dozen AGNs. Some of the notable projects are: (1) Individual
monitoring of Seyfert I galaxies (e.g., Mrk\,279, NGC\,5548,
NGC\,4151 --- Maoz et al. 1991 --- and many more by the ``AGN
Watch''\footnote{http://www.astronomy.ohio-state.edu/$\sim$agnwatch/}
projects --- Peterson 1999). (2) The Lover of Active Galaxies
(LAG) campaign (e.g., Robinson 1994). (3) The Ohio State University
monitoring program (Peterson et al. 1998). (4) The Wise Observatory and
Steward Observatory 17 Palomar-Green (PG) quasars monitoring program
by Kaspi et al. (2000). For recent reviews of the subtleties of the
reverberation mapping technique see Peterson (1993), Netzer \& Peterson
(1997), Peterson (2006) and references therein. In the following
sections I will summarize the current situation and recent studies
using reverberation mapping. I will also try to point on directions
I think reverberation mapping should take in the near future.

\section{Size---Luminosity Relation}

Peterson et al. (2004) compiled all available reverberation-mapping
data, obtained up to then, and analyzed them in a uniform and
self-consistent way to improve the determination of the time lags
and their uncertainties and derived $R_{BLR}$ for all objects with
available data. Kaspi et al. (2005) used these size measurements to
study the relation between $R_{BLR}$ and the Balmer emission line,
X-ray, UV, and optical continuum luminosities. This relation is a
fundamental relation in AGNs study since both quantities ($R_{BLR}$
and $L$) are directly obtained from {\it measurements} with minimum
assumptions and models. Once this relation is determined for the
objects with reverberation-mapping data, it is used to estimate the
mass of the BH in other AGNs by using a `single-epoch measurements'
of their luminosity and the line width (e.g., Wang \& Lu 2001; Woo
\& Urry 2002; Grupe \& Mathur 2004). To study the robustness of
the correlation Kaspi et al. (2005) used data subsamples and two
different regression methods: (1) The linear regression method of
Press et al. (1992), in which a straight-line is fitted to the data
with errors in both coordinates (known as FITEXY) and follow Tremaine
et al. (2002) procedure to account for the intrinsic scatter in the
relation.  (2) The bivariate correlated errors and intrinsic scatter
(BCES) regression method of Akritas \& Bershady (1996).

\begin{figure}[t]
\centerline{\includegraphics[width=13cm]{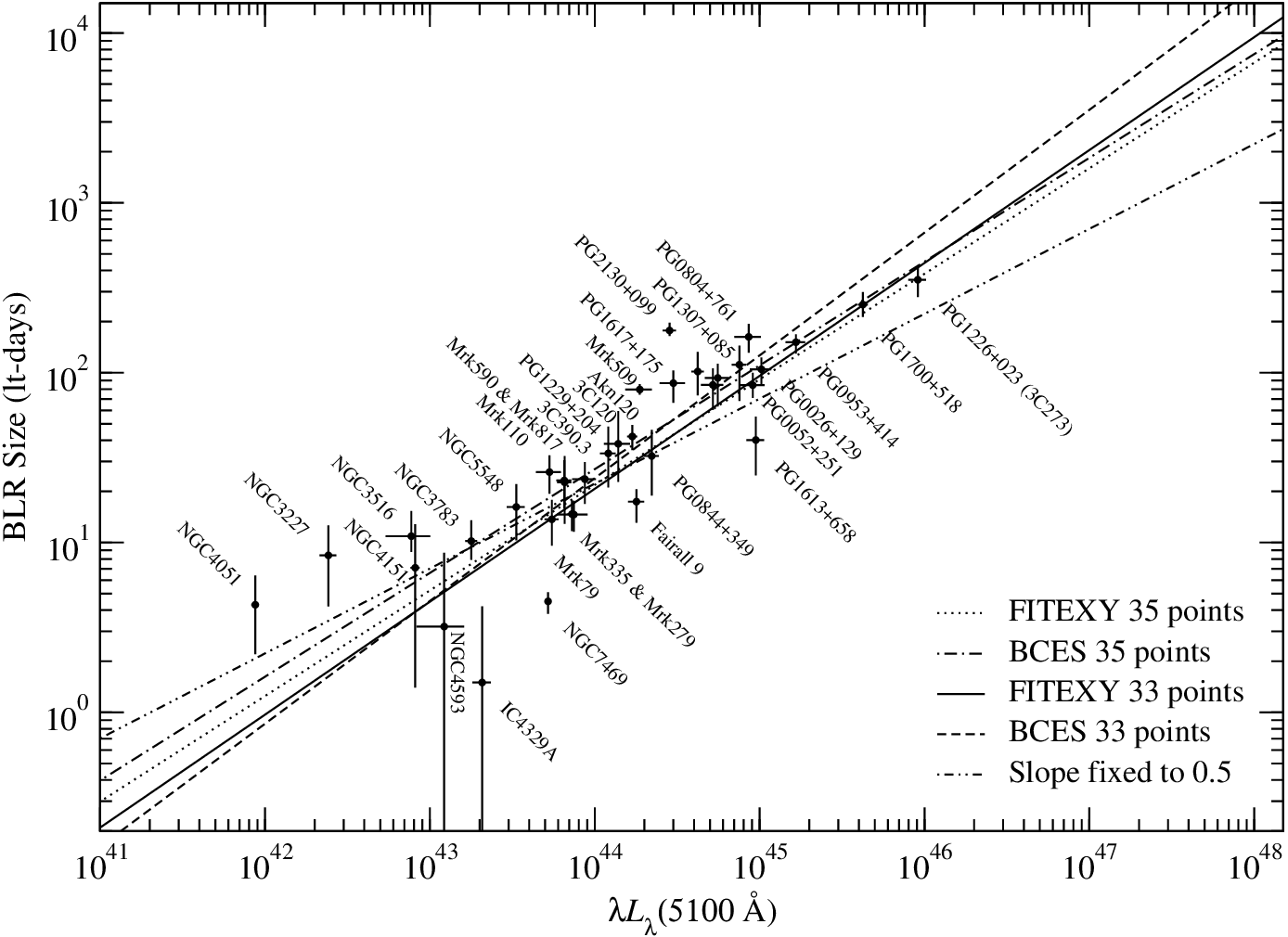}}
\caption{Balmer-line BLR size plotted versus the $\lambda
L_{\lambda}$(5100\,\AA) luminosity (in units of ergs\,s$^{-1}$). The
BLR size of each data set is determined from the averaged Balmer-line
time lags. Objects with multiple data sets have been averaged to one
point per object. See Kaspi et al. (2005) for further details. }
\end{figure}

Assuming a power-law relation $R_{BLR} \propto L^\alpha$ Kaspi
et al. (2005) find that the mean best-fitting $\alpha$ is about
$0.67\pm0.05$ for the optical continuum and the broad H$\beta$
luminosity, about $0.56\pm0.05$ for the UV continuum luminosity,
and about $0.70\pm0.14$ for the X-ray luminosity. They also find an
intrinsic scatter of $\sim 40$\% in these relations. In Fig.~1 the
mean Balmer-line BLR size versus the $\lambda L_{\lambda}$(5100\,\AA )
luminosity is plotted, with one averaged point per object. Four fits
are shown: using all 35 points, excluding the two low luminosity AGNs
(which are heavily influenced by intrinsic reddening), and with the
two fitting methods. Within the luminosity range of the measurements
($10^{43}$--$10^{46}$ ergs\,s$^{-1}$) all fits are consistent with each
other and all are well within the scatter of the points in the plot.
The disagreement of these results with the theoretical expected
slope of 0.5 indicates that the simple assumption of all AGNs having
on average same ionization parameter, BLR density, column density,
and ionizing spectral energy distribution, is not valid and there
is likely some evolution of a few of these characteristics along the
luminosity scale.

Bentz et al. (2006) used high-resolution images of the central region
of 14 of the reverberation-mapped AGNs (mostly the low luminosity
ones) and accounted for the host-galaxy star light contamination
of the AGN luminosities. Removing the star-light contribution and
excluding several points (some of which do not have measured star-light
contributions, do not have reliable H$\beta$ BLR size measurement,
or have nuclear structure and reddening that influence the luminosity
measurement), they find the power-law slope of the size---luminosity
relation to be $0.518\pm0.039$ (see Bentz' contribution in these
proceedings).

\section{Expanding the Luminosity Range}

Current reverberation mapping studies cover the luminosity range
of $\sim10^{42}$--$10^{46}$\,erg\,s$^{-1}$. Since the full AGN
luminosity range is 4 orders of magnitude larger than this range
and span the range $10^{40}$--$10^{48}$\,ergs\,s$^{-1}$, there is an
essential need to carry out reverberation mapping studies for lower-
and higher-luminosity AGNs. Hopefully such broadening of the luminosity
range will help to define better the slope of the size---luminosity relation.

Since most reverberation-mapping studies mentioned above were
based on Balmer emission lines, generally H$\beta$, and
on optical luminosity, single-epoch estimates for objects
at redshifts $z\ga 0.6$ have had to rely either on IR observations
(e.g., Shemmer et al. 2004) or on attempts to extend the optically
based size---luminosity relation to UV luminosities and UV broad
emission lines (e.g., McLure \& Jarvis 2002; Vestergaard \& Peterson
2006). While important progress has been made, there are still a number
of potential problems that need to be addressed (e.g., Maoz 2002;
Baskin \& Laor 2005). Thus, the single-epoch measurements depend on
the untested assumption that these extrapolations are valid. Although
{\it a posteriori} explanations of the physical plausibility of the
observed relations can be found, it is quite possible that subtle or
strong deviations from the relations occur at high luminosities or
redshifts (Netzer 2003).

\subsection{Reverberation Mapping of High-Luminosity AGNs}

Reverberation mapping of high-luminosity quasars is an ambitious
task. Quasars of the highest luminosities (with bolometric luminosity,
$L_{\rm bol} \approx 10^{47}$--$10^{48}$ ergs\,s$^{-1}$) are expected
to harbor some of the most massive BHs known, with $M_{\rm
BH} \ga 10^9 M_{\sun}$. More massive BHs may have slower
continuum flux variations with smaller amplitudes (e.g., Giveon et
al. 1999; Vanden Berk et al. 2004). The required observing periods
of high-luminosity quasars are also significantly lengthened by
cosmological time dilation, since such sources are typically found
at high redshifts ($z \ga 1$). On the other hand, the ability to
monitor high-$z$ objects in the rest-frame UV, in which AGN variability
amplitudes are routinely higher than in the optical, can lead to better
characterized continuum light curves. The smaller intrinsic variability
amplitude of the continuum could result in smaller flux-variability
amplitudes for the emission lines, affecting the ability to detect the
time delay in the BLR response. Furthermore, high-redshift sources
are fainter and hence more difficult to observe.  Probably due to
all of these possible problems, no reverberation measurements exist
for AGNs with $L\ga 10^{46}$\,ergs\,s$^{-1}$, and several attempts
at such measurements have so far not been successful (e.g., Welsh et
al. 2000; Trevese et al. 2006; A. Marconi 2005, private communication).

In view of the many unknowns and the opposing effects entering the
above discussion, and considering the importance of the subject,
over a decade ago, we began a reverberation-mapping program aimed at
high-luminosity, high-redshift AGNs (Kaspi et al. 2007). The sample of
11 high-luminosity quasars was selected in 1994 from the Veron-Cetty
\& Veron (1993) catalog. These are high declination ($\delta \geq
60^\circ$) objects, with observed magnitude $V\la18$, redshifts in the
range $2<z<3.4$, and in the luminosity range of $10^{46.4}\la\lambda
L_\lambda(5100\,{\rm \AA})\la10^{47.6}$ ergs\,s$^{-1}$. This is an
order of magnitude higher than other AGNs with existing reverberation
measurements (see Fig.~1).

All 11 AGNs are monitored photometrically at the Wise Observatory
(WO) 1\,m telescope since 1995 in $B$ and $R$ bands. Since the
targets have high declinations, they can be observed from the WO
for about 10 months a year, with observations scheduled about once
every month. Spectrophotometric monitoring of six of the 11 quasars
has been carried out since 1999 at the 9\,m Hobby-Eberly Telescope
(HET). Observations are carried out using a comparison star which is
observed in the slit simultaneously with the quasar and serves for
spectrophotometric calibration of the quasar under non-photometric
conditions. For details of the observational and reduction technique
see, e.g., Maoz et al. (1994) and Kaspi et al. (2000).

The continuum light curves of all 11 quasars show variations
of 10--70\% measured relative to the minimum flux. Comparing
the variability characteristics of this sample to these of the
lower-redshift PG quasar sample (Kaspi et al. 2000) the variability
of the later is about double the variability of the high-luminosity
quasars. The lower rest-frame variability measured in the continuum
for the current sample is probably a manifestation of the general
trend that high-luminosity AGNs have longer variability timescales
(e.g., Vanden Berk et al. 2004), perhaps as a result of their higher
BH masses.

None of the four Ly$\alpha$ light curves shows significant
variability. In contrast, the two C\,{\sc iii}]$\lambda$1909
light curves and all six C\,{\sc iv}$\lambda$1550 light curves
show significant variability. The variability measures of these
emission-line light curves are comparable to, or even greater than,
those of their corresponding continuum light curves.  There are
few previous AGN UV data sets with which to compare these possible
trends. The only quasar with UV variability data of similar quality is
3C273. Interestingly, Ulrich et al. (1993) noted the non-variability
of Ly$\alpha$ in this object, at a level of $<5$\%, over a period of
15 years, despite factor-of-two variations in the continuum during the
same period (there are no data for the C\,{\sc iv} and C\,{\sc iii}]
lines during that time).

\begin{figure}[t]
\centerline{\includegraphics[width=13cm]{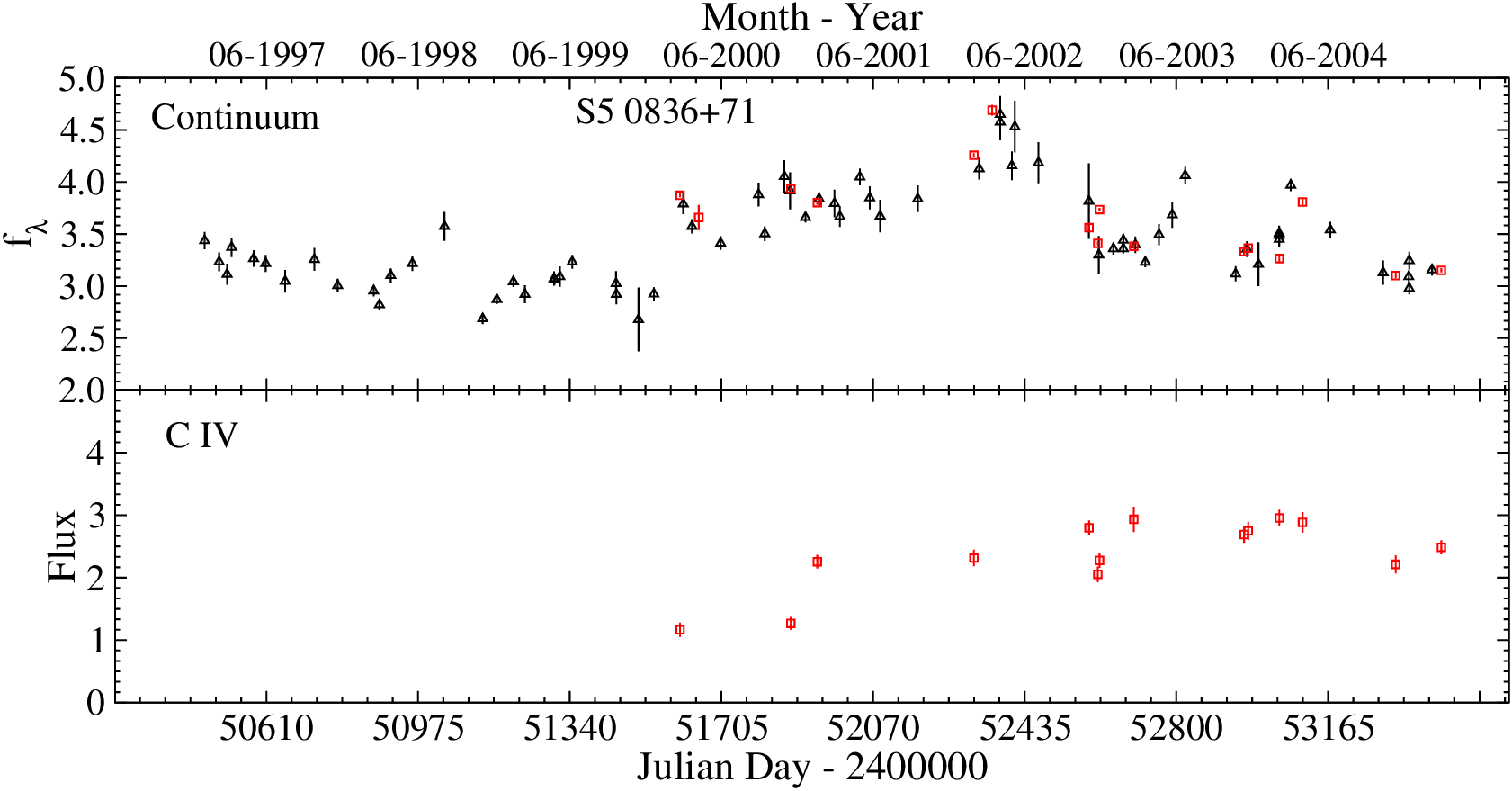}}
\caption{Optical continuum and C\,{\sc iv} light curves
for S5\,0836+71. Squares are spectrophotometric data from
the HET. Triangles are photometric data from WO. Time is
given in Julian Day ({\it bottom}) and UT date ({\it top}).
Continuum flux densities, $f_\lambda$, are given in units of
$10^{-16}$\,ergs\,cm$^{-2}$\,s$^{-1}$\,\AA$^{-1}$\ and emission-line
fluxes are given in units of $10^{-14}$\,ergs\,cm$^{-2}$\,s$^{-1}$.}
\end{figure}

The main objective of our program is to detect and measure a time delay
between the continuum and the line-flux variations in high-luminosity
AGNs. The significant continuum and line variations that was observed
during a decade demonstrate that, at least in principle, such a
measurement may be feasible. Examining the light curves of the six
quasars with emission-line data at the current stage of our project,
all but one currently suffer from either low variability amplitude
in the emission-line light curves or monotonically increasing or
decreasing continuum light curves. The one current exception is
S5\,0836+71; although the data for this quasar are still not ideal
for reverberation mapping they do allow a preliminary measurement
of the emission-line to continuum lag. Figure~2 shows the continuum
and C\,{\sc iv} emission line light curves for this object. These
light curves has largest variation among all our monitored quasars.
Figure~3 shows the CCFs for these two light curves.  The tentative
time lag between the C\,{\sc iv} line and the continuum of S5\,0836+71
is found to be $595^{+86}_{-110}$ days, or $188^{+27}_{-37}$ days in
the quasar rest frame.

\begin{figure}[t]
\centerline{\includegraphics[width=8.0cm]{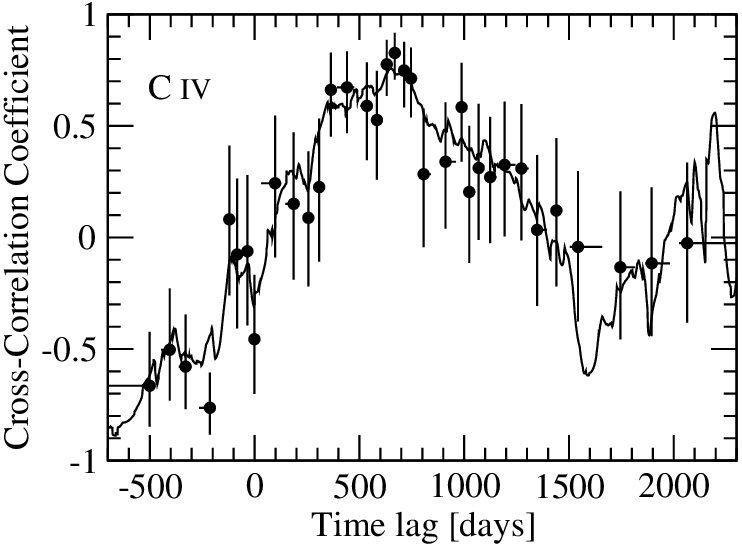}}
\caption{Cross-correlation functions, ICCF ({\it solid curve}, White
\& Peterson 1994; Gaskell 1994) and ZDCF ({\it circles with error
bars}, Alexander 1997), between the continuum and the C\,{\sc iv}
emission-line light curves of S5\,0836+71 from Fig.~2.}
\end{figure}

The mean FWHM of the C\,{\sc iv} line measured from the mean spectrum
of S5\,0836+71 is about 9700\,km\,s$^{-1}$. Using Eq.~5 of Kaspi
et al. (2000) and the time lag of 188 days, the central
mass of S5\,0836+71 is estimated to be $\sim 2.6\times10^{9}$M$_{\sun}$. This
is the highest mass directly measured for a BH in
an AGN using reverberation mapping. 3C\,273 (=PG\,1226+023),
the quasar with the highest directly measured mass so far, has a
mass of $8.9\times10^{8}$M$_{\sun}$, $\lambda L_\lambda$(1350\,\AA
)=2.0$\times10^{46}$ ergs\,s$^{-1}$, and $\lambda L_\lambda$(5100\,\AA
)=9.1$\times10^{45}$ ergs\,s$^{-1}$. Thus, S5\,0836+71 has a factor
three higher mass and a factor $\sim 6$ higher luminosity than 3C\,273.

\subsection{Reverberation Mapping of low-Luminosity AGNs}

Reverberation mapping of low-luminosity AGNs might be considered
fairly easy due to the short timescales involved and the expected
high-amplitude variability. However, as these objects are of low
luminosity there will be a need of a 3--10\,m class telescope to carry
out reverberation mapping campaigns for such objects (for example the
candidate AGNs with intermediate-mass BH from the sample by Green \&
Ho 2004). So far no reverberation mapping campaign for AGNs with optical
luminosity $\la 10^{42}$\,ergs\,s$^{-1}$ were carried out successfully.

The one exception is NGC\,4395 ($\lambda L_\lambda$(5100\,\AA
)=5.9$\times 10^{39}$\,ergs\,s$^{-1}$) in which Peterson et al. (2005)
measured the BLR size of its C\,{\sc iv} emission line to be $1\pm 0.3$
light hr. This is consistent with the size expected from extrapolating
the $R_{\rm BLR}$--$L$ relation to lower luminosities. However, two
optical campaigns to determine the H$\beta$ time lag of NGC\,4395
were so far unsuccessful due to bad weather (Desroches et al. 2006 and
contribution in these proceedings; Kaspi et al. 2007, in preparation).

Until recently, only four AGNs had measured C\,{\sc iv} reverberation
time lags: NGC\,3783, NGC\,5548, NGC\,7469, and 3C\,390.3 (see Peterson
et al. 2004, for a summary). NGC\,4395 is four orders of magnitude
lower in luminosity than those four AGNs and S5\,0836+71 is 3 orders of
magnitude higher. Thus, a preliminary C\,{\sc iv}-size---UV-luminosity
relation over 7 orders of magnitude in luminosity can be determined.
Figure~4 shows the data for the above 6 objects with the best fit
slope using the two different fitting methods; $0.551\pm0.053$ and
$0.559\pm0.025$ for the FITEXY and BCES methods, respectively.

\begin{figure}
\centerline{\includegraphics[width=10.2cm]{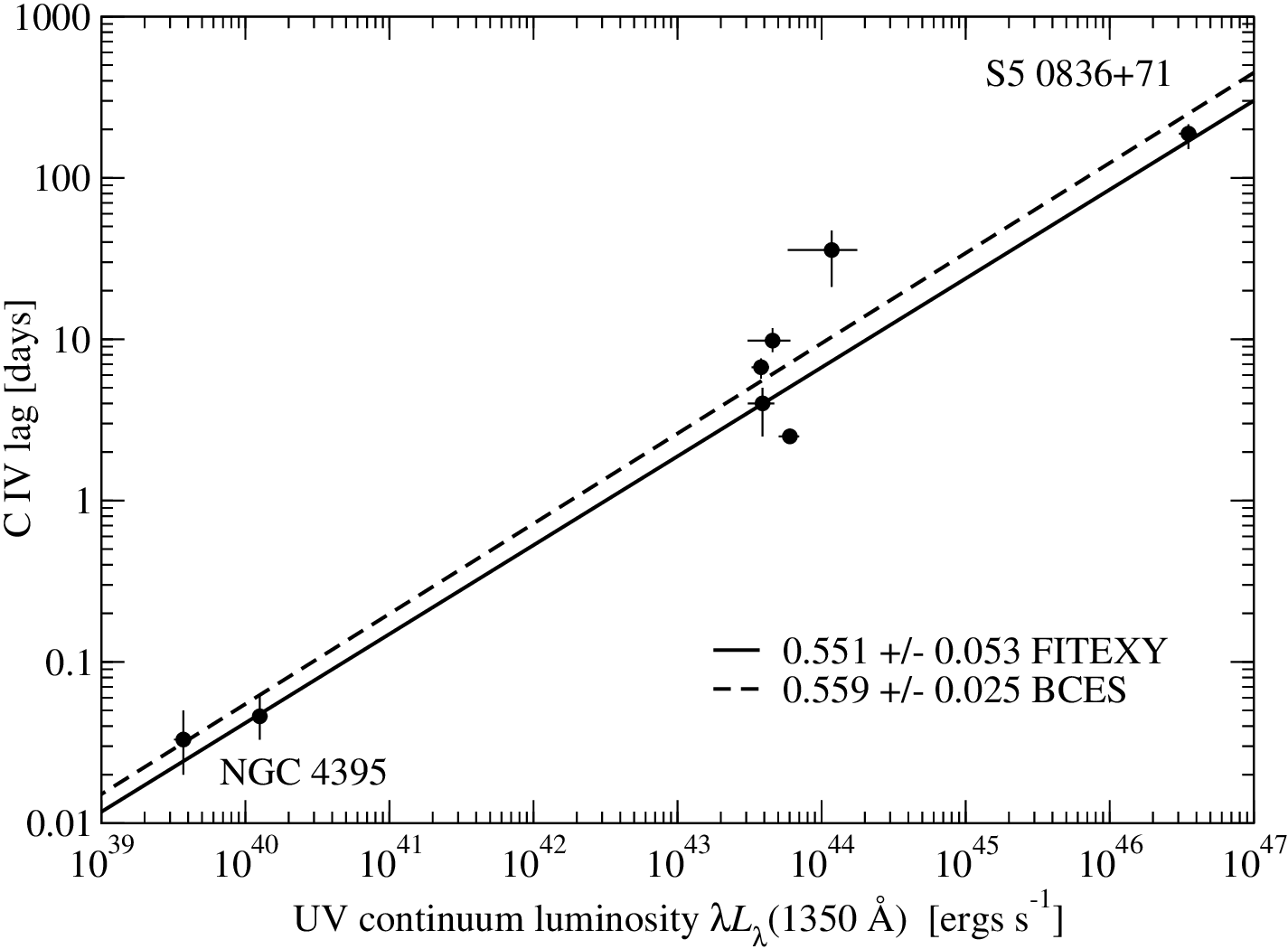}}
\caption{Size-luminosity relationship based on the C\,{\sc iv}
emission line and the UV continuum. Linear fits to the
data are shown.}
\end{figure}

\section{Further Prospects of Reverberation Mapping}

\subsection{Two-Dimension Reverberation Mapping}

The initial goal of the reverberation-mapping technique (Blandford
\& McKee 1982) is to study the geometry and kinematics of the BLR.
This was not yet achieved due to poor spectroscopic data and
insufficient time sampling. More detailed information on BLR
geometry and kinematics in AGN can be obtained by studying line
profile variations. Various researchers computed two-dimensional echo
images for specific gas motions in the BLR: outward moving gas clouds,
inward falling clouds, circulating gas clouds in a plane, or clouds
orbiting in randomly-inclined orbits (e.g., Welsh \& Horne 1991;
Perez et al. 1992; Horne et al. 2004).

Only few studies attempted the two-dimensional reverberation mapping
on actual data and none resulted with conclusive results. Recently,
Kollatschny (2003; and contribution in these proceedings) used the
monitoring data of Mrk\,110 to study the variations in the line
profiles and produced time-delay versus velocity maps. These maps
resemble a disk transfer-function maps. Kollatschny (2003) finds that
the outer line wing respond before the inner line profile and confirms
the stratification of the ionizing structure in the BLR by showing
that lower ionization lines respond after the higher ionization lines.
Thus, With recent improved sensitivity of optical telescopes and the
accumulated experience from previous mapping campaigns, it is becoming
possible to obtain the crucial information about the geometry and
kinematics of the BLR gas.

\subsection{Dust Reverberation Mapping}

IR emission in AGNs is considered to be from the alleged torus
region which is at distances larger than the BLR from the BH. Thus,
IR reverberation mapping might reveal the distance of the torus from
the BH. Only few IR monitoring campaigns were carried out in the past
two decades (e.g., Clavel et al. 1989; Glass 1992; Sitko et
al. 1993). Recently, Suganuma et al. (2006; and see contribution
in these proceedings) monitored the optical and IR emission in four
additional objects and determined time lags. Together with previous
results they are able to construct the torus-size---luminosity relation
for 10 objects. They find the torus size to strongly correlate with
the optical luminosity (the time lag is consistent with the square
root of the luminosity) and that it weakly correlates with the mass
of the BH.

\subsection{X-ray FeK$\alpha$ Reverberation Mapping}

Several studies suggested the application of the reverberation mapping
technique to the broad 6.4 keV FeK$\alpha$ line seen in the X-ray band
(e.g., Reynolds et al. 1999). This line is considered to emerge from
the accretion disk in the very close vicinity of the BH, and using
reverberation mapping will allow the measurement of the disk's
size. So far several attempts to apply this method did not produce
significant results (e.g., Ballantyne et al. 2005 and references
therein), implying either on the complicated connection between the
broad 6.4\,keV FeK$\alpha$ line flux and the X-ray continuum, or on
the fact that the X-ray data obtained so far were not sufficient for
reverberation mapping.

\section{Summary}

Over the past two decades reverberation mapping of AGNs have yield
measurements of the BLR size in about three dozen AGNs in the luminosity
range $\sim10^{42}$--$10^{46}$\,erg\,s$^{-1}$). This enables to
establish a scaling relation between the BLR size and luminosity
in AGNs which, in turn, allows the estimate of the BH mass in AGNs. Using
reverberation mapping of different emission lines implies about the
radial ionization stratification of the BLR (higher ionized specious
emits from inner BLR), and that motion of the gas in the BLR are virial
and primarily orbital. Current BLR studies should aim at broadening the
luminosity range to all AGNs ($\sim10^{40}$--$10^{40}$\,erg\,s$^{-1}$)
and first steps toward low- and high-luminosity AGNs are being taken.
Two-dimensional reverberation mapping is a promising direction which
will produce information about the geometry and kinematics of the BLR.
Reverberation mapping in the IR enables measurement of the dusty
region in AGNs (torus) which seems to surrounds the BLR. On the other
hand reverberation mapping of the inner accretion disk, using X-ray
observations, is still to be proven feasible.

\acknowledgements 

I would like to thank the organizers for a stimulating meeting
and for inviting me to give this talk. I thank my collaborators in
the high-luminosity AGNs monitoring project, Dan~Maoz, Hagai~Netzer,
W. N. Brandt, Donald P. Schneider, and Ohad Shemmer. I gratefully
acknowledge the financial support of the Colton Foundation at Tel-Aviv
University and the Zeff Fellowship at the Technion.


\end{document}